\documentclass{amsart}
\usepackage[T1]{fontenc}
\usepackage[latin1]{inputenc}
\usepackage{geometry}


\begin{document}

\title{Permutation orbifolds and their applications}

\author{P. Bantay}

\address{Institute for Theoretical Physics, Rolland Eotvos University, Budapest.}

\email{bantay@poe.elte.hu}


\thanks{Work supported by grant OTKA T32453.}

\begin{abstract}
The theory of permutation orbifolds is reviewed and applied to the study of
symmetric product orbifolds and the congruence subgroup problem. The issue of
discrete torsion, the combinatorics of symmetric products, the Galois action
and questions related to the classification of RCFTs are also discussed.
\end{abstract}
\maketitle

\section{Introduction}

The notion of permutation orbifold had been introduced in \cite{KS}, followed
by a couple of papers investigating their properties, mostly through considerations
related to modular invariance \cite{KSF}. Much later, essentially permutation
orbifold techniques have been applied in \cite{DV3} in the study of second
quantized strings. To our knowledge the first systematic study of permutation
orbifolds appeared in \cite{BHS}, where the primary field content, the genus
one characters, the modular representation and the fusion rules have been worked
out for permutation orbifolds with cyclic twist groups. These results have been
generalized to arbitrary permutation groups in \cite{PO}\cite{PO2}, which
also gave the geometric interpretation of the results through the theory of
covering surfaces. The mathematics of the construction have been clarified in
\cite{BDM} in the framework of Vertex Operator Algebras. Permutation orbifolds
have been also investigated in relation to the Orbifold Virasoro Master Equation
\cite{ovme}.

The aim of these notes is to give a sketchy overview of the basics of permutation
orbifolds, together with applications to symmetric product orbifolds and the
proof of the congruence subgroup property of RCFTs. They are not meant to be
self contained, the relevant details may be found in the cited papers. Besides
presenting standard results like the partition function and the modular representation
of permutation orbifolds, we'll touch upon such topics as discrete torsion,
the combinatorics of symmetric products, the Galois action and the classification
of RCFTs.

\section{Permutation orbifolds}

Let's begin by recalling the most important results of \cite{PO}\cite{PO2}.
We consider a permutation group \( \Omega  \) of degree \( n \), and a rational
CFT \( \mathcal{C} \). The \( n \)-fold tensor power of \( \mathcal{C} \),
i.e. the tensor product of \( n \) identical copies, admits the permutations
in \( \Omega  \) as symmetries, consequently one may orbifoldize this tensor
power by the twist group \( \Omega  \). We call the resulting theory the \( \Omega  \)
permutation orbifold of \( \mathcal{C} \), and denote it by \( \mathcal{C}\wr \Omega  \).
The central charge of \( \mathcal{C}\wr \Omega  \) is just \( n \) times the
central charge of \( \mathcal{C} \). The wreath product notation for permutation
orbifolds reflects the following basic fact: if \( \Omega  \)\( _{1} \) and
\( \Omega  \)\( _{2} \) are two permutation groups, then the \( \Omega  \)\( _{2} \)
permutation orbifold of \( \mathcal{C}\wr \Omega _{1} \) is the same as the
\( \Omega _{1}\wr \Omega _{2} \) orbifold of \( \mathcal{C} \), i.e.

\begin{equation}
\label{ass}
\left( \mathcal{C}\wr \Omega _{1}\right) \wr \Omega _{2}=\mathcal{C}\wr \left( \Omega _{1}\wr \Omega _{2}\right) 
\end{equation}
where \( \Omega _{1}\wr \Omega _{2} \) denotes the wreath product of the permutation
groups \( \Omega _{1} \) and \( \Omega _{2} \) with the standard ( imprimitive
) action \cite{wreath}. This fundamental result lies at the heart of most of
what follows, e.g. it already enables one to enumerate the primary fields of
the permutation orbifold \( \mathcal{C}\wr \Omega  \): these are in one-to-one
correspondence with the orbits of \( \Omega  \) on the set of pairs \( \left\langle p,\phi \right\rangle  \),
where \( p \) is an \( n \)-tuple of primaries of \( \mathcal{C} \), i.e.
a map \( p:\left\{ 1,\ldots ,n\right\} \rightarrow \mathcal{I} \) (we denote
by \( \mathcal{I} \) the set of primaries of \( \mathcal{C} \)), while \( \phi  \)
is an irreducible character of the double \( \mathcal{D}(\Omega _{p}) \) (cf.
\cite{DPR},\cite{LMP}) of the stabilizer \( \Omega _{p}=\left\{ x\in \Omega \, |\, xp=p\right\}  \),
the action of \( \Omega  \) on the map \( p \) being the obvious one, and
the action of \( x\in \Omega  \) on the pair \( \left\langle p,\phi \right\rangle  \)
given simply by\begin{equation}
\label{omaction}
x\left\langle p,\phi \right\rangle =\left\langle xp,\phi ^{x}\right\rangle 
\end{equation}
where \( \phi ^{x}(y,z)=\phi (y^{x},z^{x}) \). A simple counting argument gives
the number of primaries of \( \mathcal{C}\wr \Omega  \) :\begin{equation}
\label{nrprim}
\left| \mathcal{I}^{\Omega }\right| =\frac{1}{|\Omega |}\sum _{(x,y,z)\in \Omega ^{\left\{ 3\right\} }}s^{|\mathcal{O}(x,y,z)|}
\end{equation}
where \( s=\left| \mathcal{I}\right|  \) denotes the number of primaries of
\( \mathcal{C} \), \( \Omega ^{\left\{ k\right\} } \) is the set of commuting
\( k \)-tuples from \( \Omega  \), and \( \mathcal{O}(x,y,z) \) denotes the
set of orbits on \( \left\{ 1,\ldots ,n\right\}  \) of the group generated
by the triple \( (x,y,z) \). For example, if \( \Omega =S_{3} \) is the symmetric
group on three letters (with its natural permutation action), then \[
\left| \mathcal{I}^{\Omega }\right| =\frac{s^{3}+21s^{2}+26s}{6}\]

Once the primaries have been classified, the next task is to determine their
genus one characters. In order to do this, the key ingredient is to understand
the geometric aspect of orbifoldization, namely its relation to the theory of
covering surfaces. Roughly speaking, the value in the orbifold theory of a quantity
associated to a given surface is obtained as a (weighted) sum over all \( n \)-sheeted
coverings of the surface whose monodromy belongs to \( \Omega  \), the summands
corresponding to the contributions of the twisted sectors. To illustrate this
principle, let's consider the genus one partition function \( Z(\tau ) \).
In this case the surface under consideration is a torus, and the allowed coverings
are characterized by homomorphisms from the fundamental group \( \mathbb {Z}\oplus \mathbb {Z} \)
of the torus into \( \Omega  \), i.e. by commuting pairs \( (x,y)\in \Omega ^{\left\{ 2\right\} } \)
corresponding to the images of the generators of the fundamental group. The
resulting covering surface is in general not connected, its connected components
- which are all tori by the Riemann-Hurwitz formula - being in one-to-one correspondence
with the orbits of the image of the homomorphism, i.e. the subgroup generated
by \( x \) and \( y \). Each \( \xi \in \mathcal{O}(x,y) \) may be characterized
by a triple of integers \( (\lambda _{\xi },\mu _{\xi }, \)\( \kappa _{\xi }) \)
, where \( \lambda _{\xi } \) is the common length of all \( x \) orbits contained
in \( \xi  \), \( \mu _{\xi } \) is the number of these \( x \) orbits (so
that \( |\xi |=\lambda _{\xi }\mu _{\xi } \) gives the length of the orbit
\( \xi  \)), while \( \kappa _{\xi } \) is the unique non-negative integer
less than \( \lambda _{\xi } \) such that \( y^{\mu _{\xi }}x^{-\kappa _{\xi }} \)
belongs to the stabilizer of \( \xi  \). If the modular parameter of the torus
under consideration is \( \tau  \), then the modular parameter \( \tau _{\xi } \)
of the covering torus corresponding to the orbit \( \xi  \) may be expressed
as \begin{equation}
\label{covtor}
\tau _{\xi }=\frac{\mu _{\xi }\tau +\kappa _{\xi }}{\lambda _{\xi }}
\end{equation}
According to the general recipe, the genus one partition function \( Z^{\Omega } \)
of the orbifold evaluated at \( \tau  \) equals 

\begin{equation}
\label{partfun}
Z^{\Omega }\left( \tau \right) =\frac{1}{\left| \Omega \right| }\sum _{\left( x,y\right) \in \Omega ^{\left\{ 2\right\} }}\prod _{\xi \in \mathcal{O}(x,y)}Z\left( \tau _{\xi }\right) 
\end{equation}

In other words, the partition function of the orbifold is a sum over all coverings,
and the contribution of each covering equals the product of the partition functions
of its connected components. By similar considerations, the genus one character
of the primary \( \left\langle p,\phi \right\rangle  \) of \( \mathcal{C}\wr \Omega  \)
reads\begin{equation}
\label{char1}
\chi _{\left\langle p,\phi \right\rangle }(\tau )=\frac{1}{|\Omega _{p}|}\sum _{(x,y)\in \Omega _{p}^{\left\{ 2\right\} }}\overline{\phi }(x,y)\prod _{\xi \in \mathcal{O}(x,y)}\omega _{p(\xi )}^{-\kappa _{\xi }/\lambda _{\xi }}\chi _{p(\xi )}(\tau _{\xi })
\end{equation}
In this last formula, \( p(\xi ) \) denotes the value of the map \( p \) on
the orbit \( \xi  \), while for a primary \( q\in \mathcal{I} \) of \( \mathcal{C} \)
\begin{equation}
\label{omdef}
\omega _{q}=\exp \left( 2\pi i(\Delta _{q}-\frac{c}{24})\right) 
\end{equation}
denotes its exponentiated conformal weight, i.e. the corresponding eigenvalue
\( T_{qq} \) of the Dehn-twist \( T:\tau \mapsto \tau +1 \). 

One may also express the higher genus partition functions of \( \mathcal{C}\wr \Omega  \)
in terms of those of \( \mathcal{C} \) \cite{uni}. The result is that the
partition function \( Z^{\Omega } \) of the orbifold evaluated at the surface
\( \mathbb {H}/G \) reads \begin{equation}
\label{gpart}
Z^{\Omega }\left( \mathbb {H}/G\right) =\frac{1}{\left| \Omega \right| }\sum _{\varphi :G\rightarrow \Omega }\varepsilon (\varphi )\prod _{\xi \in \mathcal{O}\left( \varphi \right) }Z\left( \mathbb {H}/G_{\xi }\right) 
\end{equation}
where \( \mathbb {H} \) denotes the upper half-plane, \( G \) is the Fuchsian
group uniformizing the surface, the sum runs over all homomorphisms \( \varphi :G\rightarrow \Omega  \),
and we denote by \( \mathcal{O}(\varphi ) \) the set of orbits of \( \varphi (G) \)
on \( \left\{ 1,\ldots ,n\right\}  \), while \( G_{\xi } \) is the inverse
image under \( \varphi  \) of the stabilizer of the orbit \( \xi  \). We note
that Eq.(\ref{gpart}) is valid more generally, with \( \mathbb {H} \) denoting
an arbitrary surface and \( G \) a discrete subgroup of \( Aut\left( \mathbb {H}\right)  \),
so that in particular it covers the genus one case too (cf. Eq.(\ref{partfun})).

We have also included in Eq.(\ref{gpart}) complex phases \( \varepsilon (\varphi ) \),
which run under the name of discrete torsion (cf. \cite{Vafa}) in the orbifold
literature. These are determined by a 2-cocycle \( \vartheta \in Z^{2}(\Omega ) \)
via the following recipe \cite{DW}: one may use the homomorphism \( \varphi  \)
to pull back the 2-cocycle \( \vartheta  \) to a closed 2-form of \( \mathbb {H}/G \),
and the integral of this 2-form over the surface determines the value of \( \varepsilon (\varphi ) \).
There exists explicit expressions for \( \varepsilon (\varphi ) \) in terms
of \( \vartheta  \) \cite{Symprod}\cite{Aspinwall}, e.g. in the genus one
case (when the homomorphism is determined by a couple \( \left( x,y\right) \in \Omega ^{\left\{ 2\right\} } \))
one has \begin{equation}
\label{torusdt}
\varepsilon _{T}\left( x,y\right) =\frac{\vartheta \left( x,y\right) }{\vartheta \left( y,x\right) }
\end{equation}

The explicit knowledge of the genus one characters allows us to determine the
matrix elements of the modular transformations. The exponentiated conformal
weights of the primaries of \( \mathcal{C}\wr \Omega  \) read \begin{equation}
\label{cweight}
\omega _{\left\langle p,\phi \right\rangle }=\frac{1}{d_{\phi }}\sum _{x\in \Omega _{p}}\phi (x,x)\prod _{\xi \in \mathcal{O}(x)}\omega _{p(\xi )}^{\frac{1}{|\xi |}}
\end{equation}
with \( d_{\phi }=\sum _{x}\phi (x,1) \), while for the transformation \( S:\tau \mapsto \frac{-1}{\tau } \)
we have 

\begin{equation}
\label{Smat}
S_{\left\langle p,\phi \right\rangle }^{\left\langle q,\psi \right\rangle }=\frac{1}{|\Omega _{p}||\Omega _{q}|}\sum _{z\in \Omega }\sum _{x,y\in \Omega _{p}\cap \Omega _{zq}}\overline{\phi }(x,y)\overline{\psi }^{z}(y,x)\prod _{\xi \in \mathcal{O}(x,y)}\Lambda _{p(\xi )}^{zq(\xi )}\left( \frac{\kappa _{\xi }}{\lambda _{\xi }}\right) 
\end{equation}

The matrices \( \Lambda (r) \) which appear in this last formula play a fundamental
role in the theory. They are defined as follows: if \( r=\frac{k}{n} \) is
a rational number in reduced form, i.e. with \( k \) and \( n \) coprime and
\( n>0 \), then there exists integers \( x \) and \( y \) such that \( kx-ny=1 \),
i.e. such that the matrix \( m=\left( \begin{array}{cc}
k & y\\
n & x
\end{array}\right)  \) belongs to \( SL(2,\mathbb {Z}) \). If we denote by \( M_{pq} \) the matrix
element of the modular transformation \( \tau \mapsto \frac{k\tau +y}{n\tau +x} \)
between the primaries \( p \) and \( q \), then we define \begin{equation}
\label{Lambdadef}
\Lambda _{pq}(r)=\exp \left( 2\pi ir\left( \Delta _{p}-\frac{c}{24}\right) \right) M_{pq}\exp \left( 2\pi ir^{*}\left( \Delta _{q}-\frac{c}{24}\right) \right) 
\end{equation}
or symbolically \( \Lambda (r)=T^{r}MT^{r^{*}} \), where \( r^{*}=\frac{x}{n} \).
While these matrices are well defined, i.e. they do not depend on the actual
choice of the integers \( x \) and \( y \), their value depends on the choice
of a branch of the complex logarithm, which is involved in the computation of
rational powers of \( T \), and which should be kept fixed once for all. One
may show that the actual choice of this branch is irrelevant, a different choice
would simply amount to a relabeling of the primaries of the orbifold.

The matrices \( \Lambda (r) \) enjoy interesting properties, for example \( \Lambda (r+1)=\Lambda (r) \),
\( \Lambda (0)=S \) and \( \Lambda _{pq}(r^{*})=\Lambda _{qp}(r) \), the last
equality generalizing the symmetry of the matrix \( S \). For an integer \( n>0 \)
we have the explicit expression \[
\Lambda \left( \frac{1}{n}\right) =T^{-\frac{1}{n}}S^{-1}T^{-n}ST^{-\frac{1}{n}}\]

Once we know the matrix elements of the transformation \( S \), we can insert
them into Verlinde's formula to compute the fusion rules of the theory. The
resulting cumbersome expressions may be found in \cite{PO2}, we just note the
apparition of quantities called twisted dimensions, which are defined as

\begin{equation}
\label{twd}
\mathcal{D}_{g}\left( \begin{array}{ccc}
p_{1} & \ldots  & p_{N}\\
r_{1} & \ldots  & r_{N}
\end{array}\right) =\sum _{q\in \mathcal{I}}S_{0q}^{2-2g}\prod _{i=1}^{N}\frac{\Lambda _{qp_{i}}(r_{i})}{S_{0q}}
\end{equation}
where \( p_{1},\ldots ,p_{N}\in \mathcal{I} \) are primaries of \( \mathcal{C} \),
and \( r_{1},\ldots ,r_{N}\in \mathbb {Q} \) are rational numbers. Besides
being the basic building blocks of the fusion rules of permutation orbifolds,
twisted dimensions also appear as the partition functions of Seifert-manifolds
in the 3D topological field theory associated to \( \mathcal{C} \), leading
to Verlinde-like formulae for the traces of mapping classes of finite order,
as explained in \cite{MCG}.

\section{Symmetric product orbifolds}

The importance of symmetric product orbifolds had been recognized long ago,
e.g. for second quantized strings \cite{DV3},\cite{D} and matrix string theory
\cite{DVV}. From a geometric point of view, they amount to passing from a sigma
model describing string propagation on some target manifold \( X \) to the
sigma model for the Hilbert-scheme of \( X \). In other words, if the propagation
of a single string on \( X \) is described by the CFT \( C \), the propagation
of \( n \) identical strings should be described by the permutation orbifold
\( \mathcal{C}\wr S_{n} \), where \( S_{n} \) denotes the symmetric group
on \( n \) letters, i.e. we have to gauge the permutation symmetries. As we
don't want to fix the number of identical strings, what we are really interested
in is not the value \( A_{n} \) of some quantity \( A \) in the permutation
orbifold \( \mathcal{C}\wr S_{n} \) for a given \( n \), but rather the generating
function \[
\sum _{n=0}^{\infty }A_{n}p^{n}\]
 where \( p \) is a formal variable. For example, a beautiful result of \cite{DV3}
states that \begin{equation}
\label{spartfun1}
\sum _{n}p^{n}Z_{n}\left( \tau \right) =\exp \left( \sum _{n=1}^{\infty }p^{n}T_{n}Z\left( \tau \right) \right) 
\end{equation}
where \( Z\left( \tau \right)  \) is the genus one partition function of \( \mathcal{C} \),
and the \( T_{n} \)-s are the Hecke-operators (cf. \cite{Apostol}) which act
on \( Z(\tau ) \) via \begin{equation}
\label{thecke}
T_{n}Z\left( \tau \right) =\frac{1}{n}\sum _{d|n}\sum _{0\leq k<d}Z\left( \frac{n\tau }{d^{2}}+\frac{k}{d}\right) 
\end{equation}

There is a general combinatorial identity that lies at the heart of all related
computations. It reads \cite{Symprod} \begin{equation}
\label{comb}
\sum _{n=0}^{\infty }\frac{1}{n!}\sum _{\phi :G\rightarrow S_{n}}\prod _{\xi \in \mathcal{O}(\phi )}\mathcal{Z}(G_{\xi })=\exp \left( \sum _{H<G}\frac{\mathcal{Z}(H)}{\left[ G:H\right] }\right) 
\end{equation}
 where \( G \) is any finitely generated group, while \( \mathcal{Z} \) is
a function on the set of finite index subgroups of \( G \) that takes its values
in a commutative ring and is constant on conjugacy classes of subgroups. The
second summation on the lhs. runs over the homomorphisms \( \phi :G\rightarrow S_{n} \)
from \( G \) into the symmetric group \( S_{n} \). For a given \( \phi  \),
we denote by \( \mathcal{O}(\phi ) \) the orbits of the image \( \phi (G) \)
on the set \( \left\{ 1,\ldots ,n\right\}  \), and \( G_{\xi }=\left\{ x\in G\, |\, \phi (x)\xi ^{*}=\xi ^{*}\right\}  \)
is the stabilizer subgroup of any point \( \xi ^{*}\in \xi  \) of the orbit
\( \xi  \) - note that the lhs. of Eq.(\ref{comb}) is well defined, since
the stabilizers of points on the same orbit are conjugate subgroups. Finally,
\( \left[ G:H\right]  \) denotes the index of the subgroup \( H<G \), and
the \( n=0 \) term on the lhs. of Eq.(\ref{comb}) equals \( 1 \) by convention.As
for the proof of Eq.(\ref{comb}), one may first reduce it to the case when
\( G \) is free by using the one-to-one correspondence between subgroups of
a homomorphic image and subgroups that contain the kernel of the homomorphism.
Then one may proceed by induction on the rank of \( G \), noticing that in
the case when \( G \) has rank one, i.e. \( G=\mathbb {Z} \), Eq.(\ref{comb})
reduces to a well known formula for the generating function of the cycle indices
of symmetric groups. 

To understand the relevance of Eq.(\ref{comb}) to our problem, recall from
the previous section the expression Eq.(\ref{gpart}) for the partition function
of the orbifold. As subgroups of finite index of a Fuchsian group are themselves
Fuchsian, it follows that for a given Fuchsian group \( G \) the quantity \[
\mathcal{Z}\left( H\right) =p^{\left[ G:H\right] }Z\left( \mathbb {H}/H\right) \]
makes sense, and has all the desired properties for Eq.(\ref{comb}) to hold
(\( p \) is a formal variable and \( Z \) is the partition function of \( \mathcal{C} \)).
Inserting the above expression into Eq.(\ref{comb}) and using Eq.(\ref{gpart}),
we get \[
\sum _{n}p^{n}Z_{n}\left( \mathbb {H}/G\right) =\exp \left( \sum _{H<G}\frac{p^{\left[ G:H\right] }}{\left[ G:H\right] }Z\left( \mathbb {H}/H\right) \right) \]
where on the lhs. we recognize the generating function we were looking for.
The above argument may be extended to the genus one case as well, recovering
the result Eq.(\ref{spartfun1}) after evaluation of the right hand side.

This is not the end of the story, for it was pointed out by Dijkgraaf in \cite{D}
that, because \( H^{2}(S_{n})=\mathbb {Z}_{2} \) for \( n\geq 4 \), it is
possible to introduce non-trivial discrete torsion \cite{Vafa} in the above
models. Recall from the previous section that the introduction of discrete torsion
amounts to modifying Eq.(\ref{gpart}) by suitable phases \begin{equation}
\label{dt}
Z^{\varepsilon }_{n}(\mathbb {H}/G)=\frac{1}{n!}\sum _{\phi :G\rightarrow S_{n}}\varepsilon (\phi )\prod _{\xi \in \mathcal{O}(\phi )}Z(\mathbb {H}/G_{\xi })
\end{equation}

In the case at hand, i.e. discrete torsion corresponding to the non-trivial
cocycle \( \vartheta \in H^{2}(S_{n}) \) for \( n>3 \), there is a simple
closed formula for the discrete torsion coefficients on the torus, i.e. the
genus one case Eq.(\ref{torusdt}). To write it down let's introduce the quantities
\begin{equation}
\label{xpar}
\left| x\right| =\sum _{\xi \in \mathcal{O}(x)}\left( \left| \xi \right| -1\right) 
\end{equation}
 and \begin{equation}
\label{xypar}
\left| x,y\right| =\sum _{\xi \in \mathcal{O}(x,y)}\left( \left| \xi \right| -1\right) 
\end{equation}
 for arbitrary permutations \( x \) and \( y \), where as usual, we denote
by \( \mathcal{O}(x,y) \) the set of orbits of the group generated by \( x \)
and \( y \), and \( \left| \xi \right|  \) denotes the length of the orbit
\( \xi  \). Note that \( \left| x\right|  \) determines the parity of the
permutation \( x \), i.e. \( x \) is even or odd according to whether \( \left| x\right|  \)
is even or odd. The discrete torsion coefficients for the torus read \cite{Symprod}
\begin{equation}
\label{dtt1}
\varepsilon _{T}(x,y)=\left( -1\right) ^{\left( \left| x\right| -1\right) \left( \left| y\right| -1\right) +\left| x,y\right| -1}
\end{equation}
 for a pair of commuting permutations \( xy=yx \). There is an alternate form
of the discrete torsion coefficients that is more suitable for computations,
namely \begin{equation}
\label{dtt3}
\varepsilon _{T}(x,y)=\frac{(-1)^{\left| x,y\right| }}{4}\sum _{\alpha ,\beta \in \left\{ \pm 1\right\} }(1-\alpha -\beta -\alpha \beta )\alpha ^{\left| x\right| }\beta ^{\left| y\right| }
\end{equation}
 for \( xy=yx \). 

Armed with the above, we can now compute the generating functions in the presence
of discrete torsion. In contrast to the case without discrete torsion, we no
longer get an exponential, but rather a combination of four exponential expressions.
The final result reads \begin{equation}
\label{dtpart1}
\sum _{n=0}^{\infty }p^{n}Z_{n}^{\varepsilon }(\tau )=\frac{1}{4}\sum _{\alpha ,\beta \in \left\{ \pm 1\right\} }(1-\alpha -\beta -\alpha \beta )\exp \left\{ \sum _{n=1}p^{n}T_{n}^{\alpha \beta }Z(\tau )\right\} 
\end{equation}
 where for \( \alpha ,\beta \in \left\{ \pm 1\right\}  \) the operators \( T_{n}^{\alpha \beta } \)
acting on the partition function \( Z\left( \tau \right)  \) are defined as

\begin{equation}
\label{Yab}
T_{n}^{\alpha \beta }Z(\tau )=\frac{-(-\alpha \beta )^{n}}{n}\sum _{d|n}\sum _{0\leq k<d}\alpha ^{\frac{n}{d}}\beta ^{dk+d+k}Z\left( \frac{n\tau +kd}{d^{2}}\right) 
\end{equation}
 The close analogy with Eq.(\ref{spartfun1}) makes it tempting to interpret
the operators \( T_{n}^{\alpha \beta } \) as some kind of generalizations of
the usual Hecke-operators. Note that for \( n \) odd \( T_{n}^{\alpha \beta } \)
equals the usual Hecke-operator \( T_{n} \), independently of the values of
\( \alpha  \) and \( \beta  \).

\section{The congruence subgroup property}

A well known property of Rational Conformal Field Theories is that their genus
one characters \( \chi _{p} \) span a finite dimensional unitary representation
of the modular group \( \Gamma (1)\cong SL(2,\mathbb {Z}) \). In other words,
to any matrix \( m=\left( \begin{array}{cc}
a & b\\
c & d
\end{array}\right) \in \Gamma (1) \) there corresponds a unitary representation matrix \( M \) such that \begin{equation}
\label{modtrans}
\chi _{p}\left( \frac{a\tau +b}{c\tau +d}\right) =\sum _{q}M_{p}^{q}\chi _{q}\left( \tau \right) 
\end{equation}

\newcommand{\Q}{\mathbb {Q}}

\newcommand{\Z}{\mathbb {Z}}

\newcommand{\M}[3]{#1 \equiv #2 \, \, \left( \mathrm{mod}\, #3 \right) }

\newcommand{\cyc}[1]{\mathbb {Q}\left[ \zeta _{#1 }\right] }

\newcommand{\fr}[1]{\mathfrak {#1 }}

\newcommand{\id}{\mathbb {I}}

Of special interest are the matrices \( T \) and \( S \) representing \( t=\left( \begin{array}{cc}
1 & 1\\
0 & 1
\end{array}\right)  \) and \( s=\left( \begin{array}{cc}
0 & -1\\
1 & 0
\end{array}\right)  \). As \( s \) and \( t \) generate the modular group \( \Gamma (1) \), any
representation matrix \( M \) may be written in terms of \( S \) and \( T \).
It follows from the defining relations of \( \Gamma (1) \) that \( STS=T^{-1}ST^{-1} \)
and \( S^{4}=1 \). Moreover, it is known that \( S^{2} \) equals the charge
conjugation operator, i.e. \begin{equation}
\label{s2}
\left( S^{2}\right) _{p}^{q}=\delta _{p,\overline{q}}
\end{equation}
where \( \overline{q} \) denotes the charge conjugate of the primary \( q \). 

The modular representation enjoys some remarkable properties, the most important
ones being summarized in the celebrated theorem of Verlinde \cite{Ver}: 

\begin{enumerate}
\item \emph{\( T \) is diagonal of finite order.}
\item \emph{\( S \) is symmetric.}
\item \emph{The quantities \[
N_{pqr}=\sum _{s\in \mathcal{I}}\frac{S_{ps}S_{qs}S_{rs}}{S_{0s}}\]
are non-negative integers, being the dimension of suitable spaces of holomorphic
blocks. Here and in the sequel, the label \( 0 \) refers to the vacuum of the
theory.}
\end{enumerate}
Further properties of the modular representation have been conjectured over
the years, e.g. that its kernel \[
\mathcal{K}=\left\{ m\in \Gamma (1)\, |\, M_{p}^{q}=\delta _{p,q}\right\} \]
is of finite index in \( \Gamma (1) \), culminating in the following conjecture
\cite{CG}\cite{Bauer}\cite{Eholzer}.

Congruence subgroup property\emph{: The kernel \( \mathcal{K} \) is a congruence
subgroup, i.e. it contains the principal congruence subgroup \[
\Gamma \left( N\right) =\left\{ \left( \begin{array}{cc}
a & b\\
c & d
\end{array}\right) \in \Gamma \left( 1\right) \, |\, \M{a,d}{1}{N},\, \, \M{b,c}{0}{N}\right\} \]
for some \( N \). Moreover, the matrix entries \( M_{p}^{q} \) of modular
transformation matrices belong to the cyclotomic field \( \cyc{N} \), and \( N \)
equals the order of the Dehn-twist \( T \).}

The truth of this conjecture would have important implications in the study
of RCFTs, e.g. it would make available the apparatus of the theory of modular
functions. The first result about the conjecture was obtained in \cite{CG},
where it was shown that the conjecture holds if the order of the Dehn-twist
is odd. Unfortunately, this is rather atypical, e.g. for most of the Virasoro
minimal models the order of \( T \) is even. A proof valid for arbitrary RCFTs
had been presented in \cite{MK}, which relies on the theory of the Galois action
and the Orbifold Covariance Principle. Let's sketch these two main ingredients
of the proof.

The basic idea in the theory of the Galois action \cite{BG}\cite{CG1} is to
look at the field \( F \) obtained by adjoining to the rationals \( \mathbb {Q} \)
the matrix elements of all modular transformations. One may show that, as a
consequence of Verlinde's theorem, \( F \) is a finite Abelian extension of
\( \mathbb {Q} \). By the celebrated theorem of Kronecker and Weber this means
that \( F \) is a subfield of a cyclotomic field \( \mathbb {Q}\left[ \zeta _{n}\right]  \)
for some integer \( n \), where \( \zeta _{n}=\exp \left( \frac{2\pi i}{n}\right)  \)
is a primitive \( n \)-th root of unity. We'll call the conductor of \( \mathcal{C} \)
the smallest \( n \) which is divisible by the order of the Dehn-twist and
for which \( F\subseteq \cyc{n} \). It follows from the above that the elements
of the Galois group \( \mathrm{Gal}\left( F/\mathbb {Q}\right)  \) are (the
restriction to \( F \) of) the Frobenius maps \( \sigma _{l}:\cyc{n}\rightarrow \cyc{n} \)
that leave \( \Q  \) fixed, and send \( \zeta _{n} \) to \( \zeta _{n}^{l} \)
for \( l \) coprime to \( n \).

According to \cite{CG1}, we have \begin{equation}
\label{sls}
\sigma _{l}\left( S_{p}^{q}\right) =\varepsilon _{l}(q)S_{p}^{\pi _{l}q}
\end{equation}
for some permutation \( \pi _{l} \) of the primaries and some signs \( \varepsilon _{l}(p) \).
In other words, upon introducing the orthogonal monomial matrices \begin{equation}
\label{gldef}
\left( G_{l}\right) _{p}^{q}=\varepsilon _{l}(q)\delta _{p,\pi _{l}q}
\end{equation}
 and denoting by \( \sigma _{l}\left( M\right)  \) the matrix that one obtains
by applying \( \sigma _{l} \) to \( M \) element-wise, we have \begin{equation}
\label{slg}
\sigma _{l}\left( S\right) =SG_{l}=G_{l}^{-1}S
\end{equation}
Note that for \( l \) and \( m \) both coprime to the conductor \begin{eqnarray*}
\pi _{lm} & = & \pi _{l}\pi _{m}\\
G_{lm} & = & G_{l}G_{m}
\end{eqnarray*}
 The Galois action on \( T \) is even simpler, for \( T \) is diagonal, and
its eigenvalues are roots of unity, consequently \begin{equation}
\label{slt}
\sigma _{l}\left( T\right) =T^{l}
\end{equation}

We note that the above results follow directly from Verlinde's theorem by simple
number theoretic arguments, and they do not require the full strength of Verlinde,
for they are true even if we just require that the numbers \( N_{pqr} \) belong
to \( \Q  \), there is no need for their integrality nor positivity. 

The second ingredient of the proof, the Orbifold Covariance Principle states
that, because a permutation orbifold of an RCFT is itself an RCFT, any property
shared by all RCFTs should hold in all permutation orbifolds as well \cite{OCP}.
While this might seem tautological, it does lead to interesting results. For
example, we have seen that we may express the fusion rules of a permutation
orbifold in terms of quantities of the original theory. According to Verlinde's
theorem the fusion rule coefficients should be non-negative integers, but the
truth of this statement in the orbifold does not follow automatically from the
construction, rather it gives interesting arithmetic restrictions on the modular
representation.

In the case at hand, the Orbifold Covariance Principle states that the Galois
action on the \( S \)-matrix elements of the orbifold may be described via
suitable permutations \( \pi _{l} \) of the primaries and signs \( \varepsilon _{l} \).
As we can express the \( S \)-matrix elements in terms of modular matrices
of the original theory, this way we get information about the Galois action
on arbitrary modular matrices. A careful study leads to the following results
\cite{MK}:

\begin{enumerate}
\item For all \( l \) coprime to the conductor \begin{equation}
\label{gtcom}
G_{l}^{-1}TG_{l}=T^{l^{2}}
\end{equation}
and more generally \begin{equation}
\label{gcom}
G_{l}^{-1}MG_{l}=\sigma _{l}^{2}\left( M\right) 
\end{equation}
for any modular matrix \( M \). 
\item If \( m=\left( \begin{array}{cc}
a & b\\
c & d
\end{array}\right) \in \Gamma (1) \) with \( d \) coprime to the conductor, then \begin{equation}
\label{modrep1}
\sigma _{d}\left( M\right) =T^{b}S^{-1}T^{-c}\sigma _{d}\left( S\right) 
\end{equation}

\end{enumerate}
Eq.(\ref{gtcom}) had been conjectured in \cite{CG}, and was used to derive
the following results, which in turn were previously conjectured in \cite{Bauer}: 

\begin{itemize}
\item For \( l \) coprime to the conductor, \begin{equation}
\label{Gl}
\sigma _{l}\left( S\right) =T^{l}ST^{\hat{l}}ST^{l}
\end{equation}
where \( \hat{l} \) denotes the inverse of \( l \) modulo the conductor.
\item The conductor equals the order \( N \) of \( T \), and \( F=\cyc{N} \).
\end{itemize}
Eq.(\ref{gtcom}) has a further implication, which is important for the classification
of allowed modular representations: there exists a function \( N(r) \) such
that the conductor \( N \) of an RCFT with \( r \) primary fields divides
\( N(r) \). According to this result, for a given number of primary fields
one has only a finite number of consistent choices for the conductor \( N \)
and the matrix \( T \). Values of the upper bound \( N(r) \) for small values
of \( r \) are given in the following table: 

\vspace{0.375cm}
{\centering \begin{tabular}{|c|c|}
\hline 
\( r \) &
\( N(r) \) \\
\hline 
\hline 
2&
240\\
\hline 
3&
5040\\
\hline 
4&
10080\\
\hline 
5&
1441440\\
\hline 
\end{tabular}\par}
\vspace{0.375cm}

As to the second result, Eq.(\ref{modrep1}) implies that \[
\mathcal{K}\cap \Gamma _{1}\left( N\right) =\Gamma \left( N\right) \]
where \[
\Gamma _{1}\left( N\right) =\left\{ \left( \begin{array}{cc}
a & b\\
c & d
\end{array}\right) \in \Gamma \left( 1\right) \, |\, \M{a,d}{1}{N},\, \, \M{c}{0}{N}\right\} \]
in particular \( \mathcal{K} \) is a congruence subgroup of level \( N \).
Moreover, Eq.(\ref{modrep1}) leads to a simple arithmetic characterization
of the kernel which is suitable for explicit computations.

Let's summarize what we have found: the congruence subgroup property holds in
any Rational Conformal Field Theory, the kernel of the modular representation
may be described by arithmetic conditions, and the conductor is bounded by a
function of the number of primary fields. All these results follow ultimately
from Verlinde's theorem and the Orbifold Covariance Principle.

Of course, a host of questions remains open. Just to cite a few, it would be
interesting to understand the relation between the algebraic geometry of the
modular curve \( X\left( \mathcal{K}\right)  \) associated to the kernel and
standard properties of the RCFT. Another interesting point would be to clarify
the implications of the above results on the structure of the associated 3D
Topological Field Theory, especially in connection with the trace identities
of \cite{MCG}. Finally, these results pave the way for a systematic enumeration
of RCFTs, or at least of the allowed modular representations.

\section{Summary}

As we have seen, besides being interesting for their own sake, permutation orbifolds
have important application both in Conformal Field Theory and String Theory.
They provide a consistent framework for constructing new CFTs from old ones,
in which any interesting quantity may be expressed in terms of the corresponding
quantities of the original theory, thanks to the underlying geometric picture.
This fact supplemented by the Orbifold Covariance Principle leads to a powerful
tool for investigating deeper properties of CFTs. 

Of course, there are many open questions related to permutation orbifolds. For
example, one may ask for an algorithm which would recognize whether an RCFT
is a permutation orbifold, and output the twist group and the original theory.
This could have important computational applications. One may also speculate
about the possibility of twisting the construction by a 3-cocycle, in analogy
with the case of holomorphic orbifolds \cite{DVVV}\cite{DW}. The classification
of the conformally invariant boundary conditions (cf. \cite{JB},\cite{Chr})
of these models would be a rewarding task too.

\end{document}